\documentclass[twocolumn,showpacs,preprintnumbers,amsmath,amssymb,aps]{revtex4}
\usepackage{graphicx} 
\usepackage{dcolumn} 
\usepackage{bm}
\usepackage{latexsym}
\usepackage{eufrak}
\usepackage[latin1]{inputenc}
\usepackage{latexsym}
\usepackage{amsmath}
\usepackage{epsfig}
\usepackage{ifpdf}

\def\be{\begin{equation}}
\def\ee{\end{equation}}
\def\bea{\begin{eqnarray}}
\def\eea{\end{eqnarray}}
\begin{document}
\title{Kerr-like Phantom Wormhole}

\author{Galaxia Miranda\footnote{Part of the Instituto Avanzado de Cosmolog\'ia (IAC)
  collaboration http://www.iac.edu.mx/}}
\email{gmiranda@fis.cinvestav.mx}
\affiliation{Departamento
de F{\'\i}sica, Centro de Investigaci\'on y de Estudios Avanzados
del IPN,\\ A.P. 14-740, 07000 M\'exico D.F., M\'exico.}

\author{Tonatiuh Matos$^*$}
\email{tmatos@fis.cinvestav.mx}
\affiliation{Departamento
de F{\'\i}sica, Centro de Investigaci\'on y de Estudios Avanzados
del IPN,\\ A.P. 14-740, 07000 M\'exico D.F., M\'exico.}

\author{Nadiezhda Motelongo Garc\'ia$^*$}
\email{nmontelongo@fis.cinvestav.mx}
\affiliation{Departamento
de F{\'\i}sica, Centro de Investigaci\'on y de Estudios Avanzados
del IPN,\\ A.P. 14-740, 07000 M\'exico D.F., M\'exico.}

\begin{abstract}
In this work we study a Kerr-like wormhole with phantom matter as source. It has three parameters: mass, angular momentum and scalar field charge. This wormhole has a naked ring singularity, other wise it is regular everywhere. The mean feature of this wormhole is that the mouth of the throat lie on a sphere of the same radius as the ring singularity an avoids any observer to see or to reach the singularity, it behaves like an anti-horizon. We analyse the geodesics of the wormhole and find that an observer can go through the  geodesics without troubles, but the equator presents an infinity potential barrier which avoids to reach the throat. From an analysis of the Riemann tensor we obtain that the tidal forces permits the wormhole to be traversable for an observer like a human being.
\end{abstract}

\date{\today}



\maketitle

\section{Introduction}

The research on wormhole started with
Einstein trying to give a field
representation of particles \cite{ER}. The idea was further
developed by Ellis, \cite{Ellis}, where instead of particles
they try to model them as bridges between two regions of the
space-time. Later on, in the seminal Morris and Thorne's
work \cite{MT}, the idea of considering such solutions as 
connections between two separated regions of the Universe, has
attracted a lot of attention.

In the 80's, the astronomer Carl Sagan asked for help
to physicist and turned out that the solution proposed by
Ellis \cite{Ellis}, actually could be interpreted as the
identification, the union, of two different regions. 
The idea of Ellis is that no matter how
far the two regions are, or even if those two regions were in different
space-time, by means
of this gluing one could just identify them, and obtain a
solution which allows to go from one region to another.
Morris and Thorne \cite{MT} showed that these wormhole
solutions need to violate the energy conditions, such type of matter is called exotic (see \cite{Viser1}
for a detailed review on this subject). The solutions can exist
but they need to be generated by matter which apparently does not
exist. This is the reason why this solutions remained in the realm of fiction.

However, more and more evidence was building towards the
presence in our Universe of unknown types of matter and energy
which do not necessarily obey the energy conditions. Now we know that the
Universe is formed by $73 \%$ of dark energy. This new type of matter composes the majority in the
Universe and happens to be everywhere \cite{EO}. There
is now an agreement among the scientific community that matter
which violates some of the energy conditions is very plausible to
exit. Thus, the issue that the wormholes can be rejected due to
the type of matter that they need is, at least, diminished. The existence of wormholes in the universe is very interesting because they could be highways to visit stars and galaxies, otherwise it will be impossible to go enough far away to visit other worlds. Of course this fact is very speculative to this stage, but no observation can discard the existence of phantom energy. Nowdays we know that may be a combination of phantom and quintessence field could be the dominant component in the universe \cite{sahni}. The fact that phantom energy can be the source of wormholes \cite{ER}-\cite{EO} is more than exciting, and
give rise to investigate about the existence of stars made of this kind of matter.

Maybe the mayor problem faced by the wormhole solutions is their
stability. The stability problem of the bridges has been
studied since the $60's$ by Penrose \cite{Pen}, in connection to
the stability of the Cauchy horizons. The stability of
the throat of a wormhole was studied numerically by
Shinkay and Hayward \cite{shin}, where they show that the wormhole
proposed by Thorne \cite{MT} collapses possibly towards a black hole
and the throat closes,
when it is perturbed by a scalar field with
stress-energy tensor defined with the usual sign. 
They also find that the throat of the wormhole grows exponentially,
when the perturbation is due to a scalar field of the same type as
that making the wormhole, thus
showing that the solution is highly unstable (see also \cite{paco1}). In \cite{dario} it was conjectured that rotation or the magnetic field could stabilize a phantom star. The idea is that a rotating solution would have more possibilities of being stable, as well as more general static
spherically symmetric solutions than the one proposed by Thorne.

We look for wormhole solutions which are traversable,
that is, a test particle can go from one side of the throat to the
other without facing large tidal forces and in a finite time. We follow the
conjecture that the rotation can stabilize the wormhole, some rotating solutions were studied in the past, as an approximation \cite{rot} or as an exact solution of the Einstein equations \cite{dario}. Nevertheless, this last solution has two main problems, it is not asymptotically flat and must be matched with a static one and it has a spherical singularity surrounding the wormhole. In this work we study the solution obtained in \cite{phantom}, it has three parameters: mass, angular momentum and scalar charge. It has also a gauge parameter which determine the radius of the throat and a naked ring singularity, exactly the same one as the Kerr solution. We will show that it is possible to go though the throat going by the geodesics.

  This work is organized as follows. In section \ref{sec:linelement} we present
the metric and give some of its limits. We also write the energy density and specialize it to the north pole, showing that from the north pole an observer measure a finite energy density, corroborating the fact that from the north pole this wormhole is traversable. In section \ref{sec:Geodesics} we study the geodesics. We find that the geodesics are refuse by the ring singularity (see also \cite{Matos:2012gj}), the easiest way to go through the throat is by the polar geodesics which can go from one part of the throat to the other side. Here also we calculate the accelerations of a vector going through the throat by the polar geodesic and find that this acceleration is very weak.  In section \ref{sec:NEC} se study the energy conditions. In section \ref{sec:tidal} we analyze the tidal forces. Of course the tidal forces are infinity on the ring singularity. Finally we give some discussion in section \ref{sec:Conclusions}.

\section{The line element in different coordinates}\label{sec:linelement}

We consider a stationary and axially symmetric space-time possessing a time-like Killing vector field $(\xi)^{\alpha}=(\partial/\partial_{t})^{\alpha}$,
generating invariant time translations and a space-like Killing vector field $\mu^{\alpha}=(\partial/\partial\varphi)^{\alpha}$ generating invariant rotations with respect to $\varphi$. The corresponding line element can be
expressed in Boyer-Lindquist coordinates,
\begin{eqnarray}
ds^2=&&-f\left(dt+\omega\,d\varphi\right)^2 + \nonumber \\ &&\frac{1}{f}\left(\Delta\left(\frac{dl^2}{\Delta_1}+d\theta^2\right)+\Delta_1\,\sin^2\theta\,d\varphi^2\right),\label{eq:solBL}
\end{eqnarray}
where
\begin{eqnarray}
\Delta&=&\left(l-l_1\right)^2 + \left(l_0^2-l_1^2\right)\,\cos^2\theta,\label{eq:Delta}\\
 \Delta_1&=&\left(l-l_1\right)^2 + \left(l_0^2-l_1^2\right),\label{eq:Delta1}
\end{eqnarray}
The line element components $\omega$ and $f$ read
\begin{eqnarray}
 \omega&=&a\,\frac{\left(l-l_1\right)}{2\,\Delta}\,\sin^2\theta,\label{eq:omega}\\
f&=&\frac{\left(a^2+k_{1}^2\right)\,e^\lambda}{a^2+k_{1}^2\,e^{2\,\lambda}},\label{eq:f}
\end{eqnarray}
where the parameter $\lambda$ is given by
\begin{equation}
 \lambda=\frac{a^2+{k_1}^2}{2\,k_1\,\Delta}\,\cos\theta,
\label{eq:lambda}
\end{equation}
being $l_1, l_0$ parameters with units of distance, such that $g_{rr}>0$, that is
\begin{equation}
 l_{0}^2>l_{1}^2>0,\label{eq:l02>l12}
\end{equation}

\begin{figure}
  \centering
  \includegraphics[width=6cm]{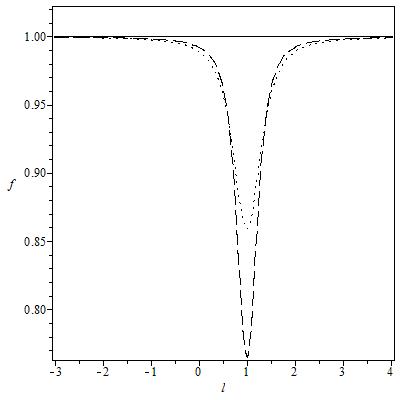}
  \caption{The function $f$ with $l_{1}=1.0$, $l_{0}=1.1$, $a=0.1$, $k=0.11$ for different values of $\theta$ ($\theta=0$ dotted line, $\theta=\Pi/2$ solid line and $\theta=\Pi/4$ dashed line). Observe that there is a minimum in 0.859 when $\theta=0$. }
\label{fig:efe}
\end{figure}

In Fig. \ref{fig:efe} we plot the function $f$ in this coordinates. Observe that the function $\lambda$ is well behave everywhere except in $\Delta=0$, just in the ring singularity (see bellow).

These parameters are related
with the size of the throat, while $a$ and $k_1$ are parameters with units of angular momentum. Line element (\ref{eq:solBL}) is a solution of the Einstein's equations $R_{\mu\nu}=-\kappa\Phi_{\mu}\Phi_{\nu}$ for an opposite sign scalar field given by
\begin{equation}
 \Phi=\frac{1}{\sqrt{2\,\kappa}}\,\lambda,
\end{equation}
where $\kappa=\frac{8\,\pi\,G}{c^4}$.

The asymptotic behaviour of metric (\ref{eq:solBL}) is as follows. For large positive or negative values of $l\rightarrow\pm\propto$, we have that
$ \omega\rightarrow 0$ and $\lambda \rightarrow 0$
thus
\[
 f \rightarrow 1
\]
and
$\Delta, \Delta_1 \rightarrow l^2$, so that the line element
\begin{equation}
ds^2 \rightarrow -dt^2 + dl^2 + l^2\left(d\theta^2 + \sin^2\theta\,d\varphi^2\right),
\end{equation}
is a flat space in spherical coordinates, $i.e.$, metric (\ref{eq:solBL}) is asymptotically flat.

\begin{figure}
 \centering
  \includegraphics[width=12cm]{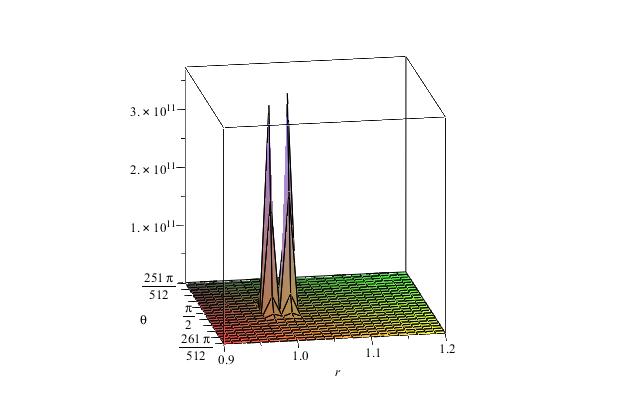}
  \caption{The Ricci scalar for the metric (\ref{eq:solBL}).}\label{fig:Ricciscalar}
\end{figure}

The scalar curvature (and the other invariants of the metric) are of the form
\begin{equation}
Invariants=\frac{F}{8\,k_{1}^2\,\Delta^{r_1}\,\Delta_1^{r_2}\,\left(a^2+k_{1}^2e^{2\,\lambda}\right)^{r_3}},
\end{equation}
where $F$ is a complicated function, which is different for the different invariants of the metric but without singularities and $r_1,\,r_2$ and $r_3$ are different coefficients depending of the invariant. In Fig. \ref{fig:Ricciscalar} we show the behaviour of the Ricci scalar of metric (\ref{eq:solBL}). The restriction (\ref{eq:l02>l12}) for the parameters $l_0$ and $l_1$ avoids that $\Delta_1$ in (\ref{eq:Delta1}) can be null, but $\Delta$ given in (\ref{eq:Delta}) is zero for $\theta=\pi/2$. Thus metric (\ref{eq:solBL}) has a naked singularity at $\Delta=0$, which represents a ring on the equatorial plane centred in the centre of the wormhole.

The ADM mass and angular momentum \cite{Kram} are given by
\begin{eqnarray}
M&=&-\frac{1}{8\,\pi}\rm{lim}_{S_t \rightarrow \infty}\oint_{S_t} \left(k-k_0\right)\sqrt{\sigma}d^2\theta \\
J&=&-\frac{1}{8\,\pi}\rm{lim}_{S_t \rightarrow \infty}\oint_{S_t} \left(K_{ij}-K\,\gamma_{ij}\right)\phi^i\,r^j\sqrt{\sigma}d^2\theta,
\end{eqnarray}
In our case we obtain
\begin{eqnarray}
M&=&-l_1\\
J&=&a
\end{eqnarray}
Thus the wormhole has a negative mass given by the size of the throat and an angular momentum given by the parameter $a$.

If we drop out the acceleration $a=0$, the rotation $\omega=0$, the metric (\ref{eq:solBL}) tranforms into
\begin{eqnarray}
ds^2&=&-f\,dt^2  \nonumber \\ &+&\frac{1}{f}\left(\Delta\left(\frac{dl^2}{\Delta_1}+d\theta^2\right)+\Delta_1\,\sin^2\theta\,d\varphi^2\right),\label{eq:solBLa0}
\end{eqnarray}
where the line element component $f$ now reads
\begin{eqnarray}
f&=&e^{-\lambda},\label{eq:fa0}
\end{eqnarray}
and the parameter $\lambda$ is give by
\begin{equation}
 \lambda=\frac{k_1}{2\,\Delta}\,\cos\theta,
\label{eq:lambda}
\end{equation}

Other interesting case is when the mass parameter $l_1=0$. The metric transforms into
\begin{eqnarray}
ds^2&=&-f\,(dt+\omega\,d\varphi)^2 +\frac{1}{f}\left(\frac{l^2 + l_{0}^{2}\,\cos^2\theta}{l^2 + l_{0}^2}dl^2\right.  \nonumber \\ &+&\left.(l^2 + l_{0}^2)\,\left[\frac{l^2 + l_{0}^2\,\cos^2\theta}{l^2 + l_{0}^2}d\theta^2+\sin^2\theta\,d\varphi^2\right]\right),\nonumber\\
\label{eq:solBLa0}
\end{eqnarray}

Observe that the term inside of the bracket $[\,\, ]$ is the solid angle element, but with a factor multiplying $d\theta$. This is the modification of the solid angle due to the axial symmetry.

\begin{figure}[htp]
  \centering
  \includegraphics[width=6cm]{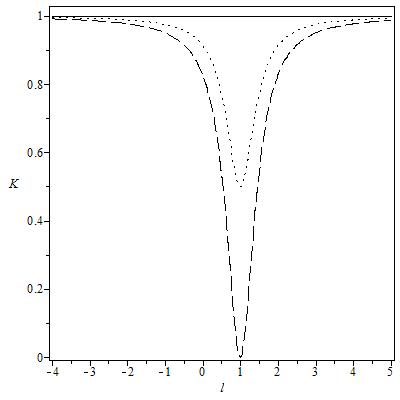}
  \caption{The function $K$ with $l_{1}=1$, $l_{0}=1.1$, $a=0.1$, $k=0.11$ for different values of $\theta$. Observe that $K=1$ everywhere except close to the ring singularity.}\label{fig:K}
\end{figure}

To see the throat of the wormhole, it is convenient to write metric (\ref{eq:solBL}) as
\begin{eqnarray}
ds^2=&&-f\left(dt+\omega\,d\varphi\right)^2 + \nonumber \\ &&\frac{K}{f}\,dl^2+\frac{\Delta_1}{f}\,\left[K\,d\theta^2+\sin^2\theta\,d\varphi^2\right],\label{eq:solBLu2}
\end{eqnarray}
where $K=\Delta/\Delta_1$. Again, the quantity between the brackets $[\,\,]$ can be interpreted as the solid angle, but now with a modification given by the $K$ function multiplying $d\theta^2$, see the plot of $K$ for different values of $\theta$ in Fig. \ref{fig:K}. The modification of the solid angle is due the fact that the space-time is now axialy symmetric and not spherically symmetric. Nevertheless, in this case, this modification is small everywhere, except near a sphere of radius $l_1$, the radius of the ring singularity, indicating that the space-time is almost spherically symmetric outside this sphere, (see figure \ref{fig:K}). Amazingly, the space-time is just spherically symmetric at the equator ($\theta=\pi/2$) of the phantom star.

\begin{figure}[htp]
  \centering
  \includegraphics[width=6cm]{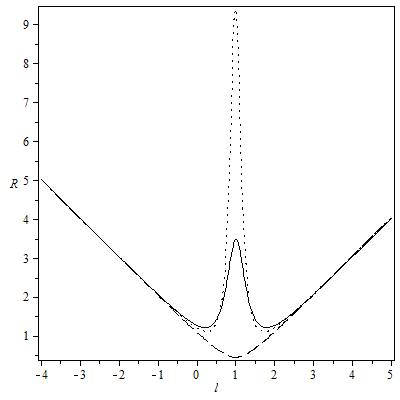}
  \caption{The function $R = \sqrt{\Delta_1/f}$ with $l_{1}=1$, $l_{0}=1.1$, $a=1$, $k=1$, for different values of $\theta$. Amazingly, the function $R$ is well behave everywhere, and is smooth on the equator ($\theta=\frac{\pi}{2}$ solid line). }
\label{fig:R(u)}
\end{figure}

The function $\Delta_1/f$ can be also interpreted as the radial coordinate to the square. In this coordinates this function never reaches the value zero, see Fig. \ref{fig:R(u)}, indicating that the area of this region is never zero and thus that there is a throat. 

Now we write metric (\ref{eq:solBL}) in cylindrical coordinates $(\hat\rho,z)$, \cite{Kram}. These coordinates are defined by the transformation,
\begin{eqnarray}
l&=&\sqrt{\hat\rho^2+z^2}-\frac{l_{0}^2-l_{1}^2}{4\,\sqrt{\hat\rho^2+z^2}} + l_{1}\nonumber\\
\theta&=&\arctan\,\left(\frac{\hat\rho}{z}\right)
\end{eqnarray}
with inverse given by
\begin{eqnarray}
\hat\rho&=&\frac{1}{2}\left(l-l_1+\sqrt{\Delta_1}\right)\,\sin\theta\nonumber\\
z&=&\frac{1}{2}\left(l-l_1+\sqrt{\Delta_1}\right)\,\cos\theta
\end{eqnarray}
leaving $\left(t,\varphi\right)$ as before. We can rewrite the line element as
\begin{eqnarray}
ds^2=&&-f\left(dt+\omega\,d\varphi\right)^2 +  \\ &&\frac{1}{\left(\hat\rho^2+z^2\right)^2\,f}\left(K\,\left(d\hat\rho^2+dz^2\right)+\hat\rho^2\,\Delta_+\,d\varphi^2\right), \nonumber
\end{eqnarray}
being
\begin{eqnarray}
\Delta_\pm&=&\hat\rho^2+z^2\pm\frac{{l_0}^2-{l_1}^2}{4}, \nonumber\\
K&=&{\Delta_-}^2 + \left({l_0}^2-{l_1}^2\right)\,z^2, \nonumber\\
\omega&=&a\frac{\hat\rho^2\,\Delta_-}{2\,\sqrt{\hat\rho^2+z^2}\,K},\nonumber\\
\lambda&=&\left(\frac{a^2+{k_1}^2}{2\,k_1}\right)\frac{\sqrt{\hat\rho^2+z^2}\,z}{K},
\end{eqnarray}
where $f$ has the same form as before, (\ref{eq:f}). Using the Weyl coordinates \cite{Kram} $(\rho,\zeta)$ defined by
\begin{eqnarray}
\rho&=&\sqrt{l^2-2\,l_1\,l+l_0^2}\,\sin(\theta)\nonumber\\
\zeta&=&(l-l_1)\,\cos(\theta)
\end{eqnarray}
we can write the metric in terms of the prolate coordinates defined as
\begin{eqnarray}
\rho&=&\sigma(x^2-1)(1-y^2)\nonumber\\
\zeta&=&\sigma\,x\,y
\end{eqnarray}
being $\sigma^2=l_0^2-l_1^2$. The inverse of this coordinates is given by
\begin{eqnarray}
 2\,\sigma\,x&=&r_++r_-\nonumber\\
2\,\sigma\,y&=&r_+-r_- \nonumber\\
r_{\pm}^2&=&\rho^2+(z\pm\sigma)^2
\end{eqnarray}

The relationship between the prolate and the Boyer-Lindquist coordinates is
\begin{eqnarray}
 \sigma\,x&=&l-l_1\nonumber\\
y&=&\cos(\theta)
\end{eqnarray}

In prolate coordinates, metric (\ref{eq:solBL}) is
\begin{eqnarray}
ds^2&=&-f\left(dt+\omega\,d\varphi\right)^2  \nonumber\\ &+&\frac{\sigma^2(x^2-y^2)}{f}\left[\frac{dx^2}{x^2-1}+\frac{dy^2}{1-y^2}\right]\nonumber\\
&+&\sigma^2(x^2-1)(1-y^2)\,d\varphi^2,\label{eq:solWeyl}
\end{eqnarray}
where the functions $\lambda$ and $\omega$ written in these coordinates read
\begin{eqnarray}
 \lambda&=&\frac{a^2+k_1^2}{2\,k_1\Delta}\,y\nonumber\\
  \omega&=&\frac{a\,\sigma\,x}{2\,\Delta}\,(1-y^2)
\end{eqnarray}
In these coordinates function $\Delta$ is
\begin{equation}
\Delta=\sigma^2\,(x^2+y^2)
\end{equation}

The wormhole geometry may be described by an embedding diagram in three dimensional Euclidean space at fixed moment in time and for a fixed value of $\theta$. Accordingly,
we will write the line element as follows
\begin{eqnarray}
ds^{2}&&=f^{-1}K\left(\frac{dl}{d\rho}\right)^{2}d\rho^{2}+\rho^{2}
\left(\sin^{2}(\theta_{0})-\frac{\omega^{2}f^{2}}{\Delta_1}\right)d\varphi^{2},\nonumber\\ 
\end{eqnarray}
where $\rho^{2}=f^{-1} \Delta_1$. The resulting surface of revolution
has the parametric form $h(\rho,\varphi)=(\rho\cos(\varphi),\rho\sin(\varphi),z(\rho,\theta_{0}))$. The requirements are met by $z=z(\rho,\theta_0)$ for any fixed $\theta$ such that:
\begin{eqnarray}
\left(\frac{dz}{d\rho}\right)^{2}=f^{-1}K\left(\frac{dl}{d\rho}\right)^{2}-1.\label{C1}
\end{eqnarray}
Furthermore, the flare-out condition $\frac{d^{2}\rho}{dz^{2}}>0$ (for the upper universe) at or near the throat, are
\begin{eqnarray}
\nonumber
&&\frac{d^2\rho}{dz^{2}}=(f^{-1}K-(\frac{d\rho}{dl})^{2})^{-1}\frac{d^2\rho}{dl^{2}}\\
&&+\frac{1}{2}\frac{d\rho}{dl}\frac{d}{dl}(f^{-1}K-(\frac{d\rho}{dl})^{2})^{-1}>0.\label{C2}
\end{eqnarray}

\begin{figure}[ht]
  \centering
  \includegraphics[width=6cm]{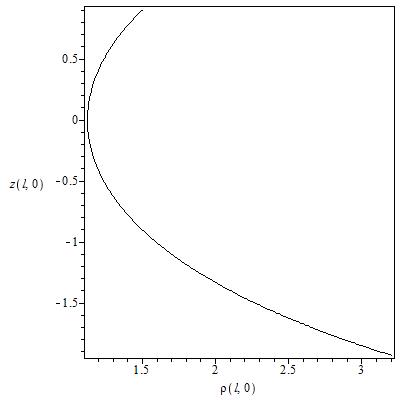}
  \caption{Function $z$ for fixed $\theta=0$. Here we set $l_1=1$, $l_2=1.5$, $a=0.1$ and $k=0.11$. This plot shows embedding diagram for a wormhole with metric given by (\ref{eq:solBL}). Since our metric is axially symmetric for different values of $\theta$ the shape of the throat may change. }\label{fig:Embendding-d}
\end{figure}

\begin{figure}[ht]
  \centering
  \includegraphics[width=6cm]{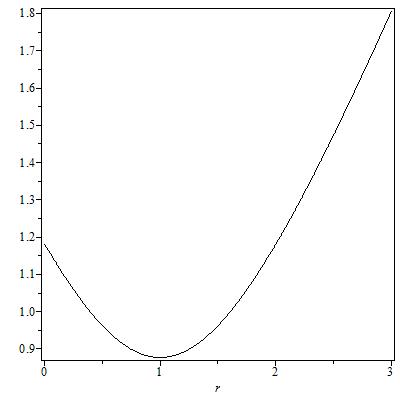}
  \caption{Function $z$ for fixed $\theta=\pi$. The throat is located at $l=1$. Here we set $l_1=1$, $l_2=1.5$, $a=0.1$ and $k=0.11$. The flare out condition is satisfied, i.e., it is positive near the throat.}\label{fig:flare-out-c}
\end{figure}

The embedding surface give by (\ref{C1}) and the flare-out condition (\ref{C2}) are plotted using the values: $l_1=1$, $l_2=1.5$, $a=0.1$ and $k=0.11$, see Fig.\ref{fig:Embendding-d} and Fig.\ref{fig:flare-out-c}. It is important to mention that given this values the flare out condition holds everywhere except in a vicinity of the singularity ($\theta=\pi/2$ and $l=1$).

\section{Geodesics} \label{sec:Geodesics}

We start with the study of the geodesics using the Boyer-Lindquist coordinates. We first are interested in radial geodesics to see whether an observer can penetrate the wormhole or not. Of course the ring singularity avoids any observed to penetrate the wormhole, at least going by the equator. Nevertheless, the easiest possibility for a traveler for going trough the troat is that the observer travels by the polar geodesic, that means, the line that joints the north pole with the south pole. In the surface of the sphere of radius $l=l_1$ the traveler meets the mouth of the throat, where the tidal forces reduce its magnitude due to the symmetry of the singularity as we saw in the latter section. In order to see that, we obtain the polar geodesics in the different coordinate systems.

For doing so, let $\tau$ be an affine parameter and  $u^\mu=(\dot t,\dot r,\dot \theta,\dot \varphi)$, with $ {\dot t}=\frac{d t}{d\tau}$, etc., the vector velocity of an observer, such that the equation $u^\mu u_\mu=-1$ holds. In Boyer-Lindquist coordinates this expression reduces to
\begin{eqnarray}
-1&=&-f\left(\dot t+\omega\,\dot\varphi\right)^2  \nonumber \\ &+&\frac{1}{f}\left(\Delta\left(\frac{\dot l^2}{\Delta_1}+\dot\theta^2\right)+\Delta_1\,\sin^2\theta\,\dot\varphi^2\right),\label{eq:LagranBL}
\end{eqnarray}
For the polar geodesic $\theta=0$, this implies that $\Delta_1=\Delta$ and $\omega=0$, that means that from this geodesic, the observer does not feel the rotation of the wormhole. The equation (\ref{eq:LagranBL}) reduces to
\begin{equation}
 -1=-f\dot t^2 +\frac{1}{f}\dot l^2,\label{eq:LagranBL2}
\end{equation}
In this metric $\frac{\partial}{\partial t}$ is a Killing vector, thus we have that $f\dot t=\sqrt{2}E$, being $E$ a constant. Then the geodesic equation (\ref{eq:LagranBL2}) reduces to
\begin{equation}
 \frac{1}{2}\dot l^2+\frac{1}{2}f=E^2\label{eq:LagranBL3}
\end{equation}

This last is a dynamical equation where we can know the motion of the test particle if we know the potential, in this case given by $1/2\,f$. 

From the geodesic equations
\begin{equation}\label{e:geodes}
\frac{d^{2}x^{\alpha}}{d\tau^{2}}+\Gamma^{\alpha}_{\beta\gamma}\frac{dx^{\beta}}{d\tau}\frac{dx^{\gamma}}{d\tau}=0
\end{equation}
where $\tau$ parametrize the curve $x^{\alpha}(\tau)$, we can obtain the same results.
In terms of the prolate coordinates, the non-vanishing Christoffel symbols of this metric are given by
\begin{equation}
\begin{array}{cc}
  \Gamma^{x}_{xx}=-\frac{f_{x}}{2f}, & \Gamma^{x}_{tt}=-\frac{ff_{x}}{2\sigma^{2}} \\
  \Gamma^{t}_{tx}=-\frac{f_{x}}{2f}, &  \Gamma^{t}_{xt}=\frac{f_{x}}{2f}.
\end{array}
\end{equation}
Thus the geodesic equations reads
\begin{equation}
\ddot{x}+\Gamma^{x}_{xx}\dot{x}^{2}+\Gamma^{x}_{tt}\dot{t}^{2}=0
\end{equation}
\begin{equation}
\ddot{t}+2\Gamma^{t}_{xt}\dot{x}\dot{t}=0
\end{equation}
Explicitly, the geodesic equation for the variable $t$ on the north pole is given by
\begin{equation}\label{e:soltemporal}
 \qquad f\dot{t}=\sqrt{2}E
\end{equation}
being $E$ the integrations constant and the corresponding one for the variable $x$
\begin{equation}\label{e:solespacial}
\frac{\sigma^{2}}{f}\dot{x}^{2}-f\dot{t}^{2}=-1
\end{equation}
Again we can write both equations (\ref{e:soltemporal}) and (\ref{e:solespacial}) together to obtain
\begin{equation}\label{e:georadial}
\frac{1}{2}\dot{x}^{2}+\frac{f}{2\sigma^{2}}=E^2.
\end{equation}
In order to have a better interpretation of equation (\ref{e:georadial}) we can transform it into
\begin{equation}
\frac{1}{2}\dot{x}^{2}-\frac{f}{2\sigma^{2}}+\frac{1}{2\sigma^{2}}=E^2+\frac{1}{2\sigma^{2}}
\end{equation}
where now the potential and the total energy are given by
\begin{equation}\label{e:newpotencial}
V\rightarrow-\frac{f}{2\sigma^{2}}+\frac{1}{2\sigma^{2}}  \qquad
E^2\rightarrow E^2+\frac{1}{2\sigma^{2}}
\end{equation}

Interpreting the dynamical equation
$\frac{\dot{x}^{2}}{2}+V=E^2$, we see that the potential $V$ is a potential well, the particle can fall into the well and remain there or, if the total energy of the test particle is bigger than $E^2$, the test particle continues through the throat. It is important to mention that this analysis is only valid for null geodesics with $\theta=0$, since other values of $\theta=const.$ lead us to non-physical geodesics.

We can integrate equation (\ref{e:georadial}) in terms of the variables $\tau$ and $x$. This equation transform into
\begin{equation}
\frac{d\tau}{dx}=\pm\left(2E^2+\frac{f}{\sigma^{2}}\right)^{-1/2}
\end{equation}
which can be reduced into quadratures as
\begin{equation}
\tau=\pm\int\left(2\,E^2- \frac{\left( {a}^{2}+{k_{{1}}}^{2} \right)
{e^{{\frac {{a} ^{2}+{k_{{1}}}^{2}}{2k_{{1}} \left(
{\sigma}^{2}{x}^{2}+{l_{{0}}}^{2}-{ l_{{1}}}^{2} \right)
}}}}}{\sigma^{2}\left( {a}^{2}+{k_{{1}}}^{2}{e^ {{\frac
{{a}^{2}+{k_{{1}}}^{2}}{k_{{1}} \left( {\sigma}^{2}{x}^{2}+{l_
{{0}}}^{2}-{l_{{1}}}^{2} \right) }}}} \right)} \right)^{-1/2}dx
\end{equation}
\begin{figure}[htp]
  \centering
 \includegraphics[width=8cm]{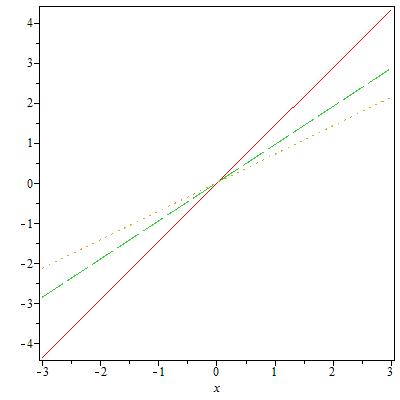}
 \caption{Time-like polar geodesics in the proper time for the values $E^2=0.5$ (solid line), $E^2=0.75$ (dashed line) and $E^2=1$ (doted line) of the total energy. Here we set $l_{1}=1$, $l_{0}=1.5$, $a=0.1$, $k_1=0.11$ and $\sigma=1$}\label{fig:temporal}
\end{figure}
This integration can be performed numerically, the results are shown in Fig. \ref{fig:temporal}, where we see that an observer can travel through the throat in finite proper time $\tau$. The plot show different possibilities for the journey.

For the general case, with $\theta\neq const.$,  in terms of momenta 
\begin{equation*}
p_l = \frac{\Delta}{\Delta_1 f} \dot{l}, \hspace{1.0cm} p_\theta = \frac{\Delta}{f} \dot{\theta},    
\end{equation*}
and the conserved quantities
\begin{equation*}
E=f (\dot{t}+\omega \dot{\varphi}), \hspace{1.0cm}  L+\omega E=\frac{f}{\Delta_1 \sin^2 \theta}\dot{\varphi}
\end{equation*}
the Euler-Lagrange equations for the geodesics of \eqref{eq:solBL} are, 
\begin{eqnarray}
\dot{p_l} &=& \frac{(a^2+k_1^2)(l-l_1)\cos \theta}{2 k_1 f_- \Delta^2}E^2 
			+\frac{a (l-l_1) f \Delta_-}{\Delta_1 \Delta^2} E(L+\omega E) \nonumber \\
			&&+\frac{(l-l_1) f^2 \Delta_1}{\Delta^2} \left(\frac{1}{f}+\frac{(a^2-k_1^2) \cos \theta}{2 k_1 \Delta f_-}\right) \left(p_l^2+\frac{p_\theta^2}{\Delta_1} \right) \nonumber \\
			&&+\frac{f^2}{\Delta_1^2 \sin^2 \theta} \left(\frac{1}{f}+\frac{(a^2-k_1^2)(l-l_1)\Delta_1 \cos \theta}{2 k_1 \Delta^2 f_-}\right)(L+\omega E)^2 \nonumber \\
			&&- \frac{(l-l_1)f}{\Delta} p_l^2, \label{PL}
\end{eqnarray}
\begin{eqnarray}
\dot{p_\theta} &=& \frac{(a^2-k_1^2)\sin \theta \Delta_-}{4 k_1 \Delta^2 f_-} E^2
				-\frac{a(l-l_1)\sin 2\theta f}{\Delta^2 \sin^2 \theta}E(L+\omega E) \nonumber \\
				&&+\frac{f^2}{\Delta_1 \sin \theta} \left(\frac{\sin 2\theta}{2 f \sin^3 \theta} 
				+\frac{(a^2-k_1^2) \Delta_-}{4 k_1 \Delta^2 f_-}\right)(L+\omega E)^2  \nonumber \\
				&&+\frac{f^2}{K \Delta}\left(\frac{(a^2-k_1^2) \Delta_- \sin \theta}{4 k_1 \Delta f_-}
				-\frac{(l_0^2 - l_1^2) \sin 2\theta}{2 f}\right) \nonumber \\
				 &&\times \left(p_l^2+\frac{p_\theta^2}{\Delta_1}\right). \label{PT}
\end{eqnarray}
being \[\Delta_- = (l-l_1)^2-(l_0^2-l_1^2)\] and \[f = \frac{(a^2-k_1^2)e^{\lambda}}{a^2-k_1^2 e^{2 \lambda}}.\]

\begin{figure}[ht]
  \includegraphics[width=8cm]{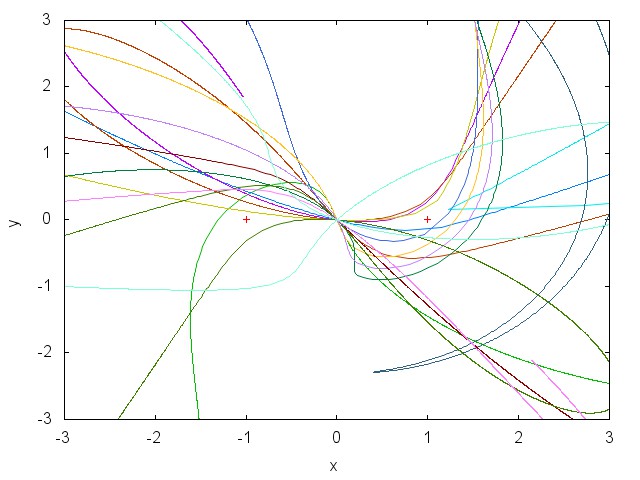}
  \caption{In this plot we show the geodesics of the wormhole falling from different values of $\theta$, with the values $l_{1}=1$, $l_{0}=1.1$, $a=0.1$, $k=0.11$, $E=1.0$ and $L=0.5$. With the singularity (the crosses) at $l=l_1$ and $\theta=\pi/2$.}\label{fig:geodd}
\end{figure}

This system can be integrated analytically for $\theta=0$. For the general case with for a non constant $\theta$ it can be solved numerically. In fig. \ref{fig:geodd} we can see the geodesics for different initial values of $\theta$. As we can see no null-geodesic is able to reach the singularity, even when the geodesic falls into the wormhole (see \cite{Matos:2012gj}).

\section{NEC Violation}\label{sec:NEC}
To make the analysis tractable, we choose an orthonormal basis
$e_{\hat{t}},e_{\hat{r}},e_{\hat{\theta}},e_{\hat{\phi}}$,

\begin{eqnarray}
\nonumber
&&e_{\hat{t}}=f^{-\frac{1}{2}}e_{t},\nonumber  \\
&&e_{\hat{r}}=(f^{-1}K)^{\frac{1}{2}}e_{r}, \nonumber \\
&&e_{\hat{\theta}}=f^{-1}K\Delta_1^{-\frac{1}{2}}e_{\theta},\nonumber \\
&&e_{\hat{\varphi}}=\frac{\omega e_{t}+e_{\varphi}}{\sqrt{f^{-1}K \Delta_1}\sin(\theta)}. \label{tetrad}
\end{eqnarray}

An important condition to build wormholes is the violation of the Null Energy Condition (NEC) at or near the throat, i.e., $T_{\hat{\alpha}\hat{\beta}}\mu^{\hat\alpha}\mu^{\hat\beta}<0$, for all null vector $\mu$.
For simplicity we use a radial outgoing null vector, $\mu=e_{\hat{t}}+e_{\hat{r}}$, thus
\begin{eqnarray}
T_{\hat{\alpha}\hat{\beta}}\mu^{\hat{\alpha}}\mu^{\hat{\beta}}=
T_{\hat{t}\hat{t}}+T_{\hat{r}\hat{r}}=\frac{1}{8\pi}(R_{\hat{t}\hat{t}}+R_{\hat{r}\hat{r}}).
\label{NEC}
\end{eqnarray}

\begin{figure}[ht]
  \centering
  \includegraphics[width=8cm]{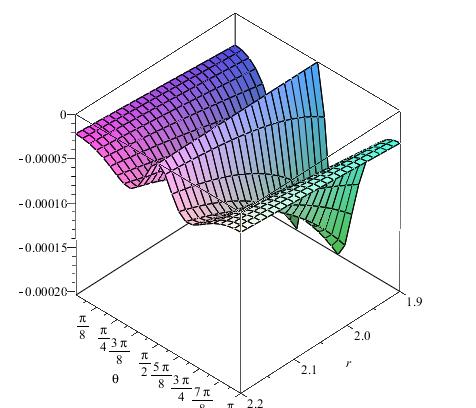}
  \caption{$T_{\hat{\alpha}\hat{\beta}}\mu^{\hat\alpha}\mu^{\hat\beta}$ for  $\mu=e_{\hat{t}}+e_{\hat{r}}$ for the tetrad \eqref{tetrad}. Here we set $l_1=1$, $l_2=1.5$, $a=0.1$ and $k=0.11$. This plot shows the violation of the NEC.}\label{fig:NEC}
\end{figure}

The expression associated to (\ref{NEC}) is very large and it is not interesting to write it down. Instead of that we plot the expression at and near the throat, see Fig. \ref{fig:NEC}. We see that, as expected, the NEC is violated in this region.

\section{The Gravitational Tidal Forces}\label{sec:tidal}

From the physical point of view it is important to know whether the wormhole is traversable or not. In this section we study the tidal forces constraints \cite{MT},\cite{rot} of metric \eqref{eq:solBL}.
For simplicity we consider a traveller going radially through the wormhole, beginning at rest in space station in the lower universe and ending at rest in space station in the upper universe. We introduce the orthonormal basis of its own reference frame,
\begin{eqnarray}
\nonumber
&&e_{\hat{0}}=\gamma e_{\hat{t}}\mp \gamma(v/c)e_{\hat{r}},\, \\ \nonumber
&&e_{\hat{1}}=\mp\gamma e_{\hat{r}}+\gamma(v/c)e_{\hat{t}}, \\ 
&&e_{\hat{2}}=e_{\hat{\theta}}, \hspace{0.5cm} e_{\hat{3}}=e_{\hat{\varphi}}.
\end{eqnarray}
being $\gamma=[1-(\frac{v}{c})^2]^{-\frac{1}{2}}$. For simplicity we assume here that the traveller do not feel an acceleration larger than about 1 Earth's gravity.
Following the same idea as in the work of Morris and Thorne \cite{MT} and \cite{rot}, the radial tidal constraint is given by
\begin{eqnarray}
|R_{\hat{1}\hat{0}\hat{1}\hat{0}}|\leq g_{\oplus}/c^{2}\times 2m\approx 1/(10^{5}km)^{2},
\end{eqnarray}
where the height of our traveller is $2m$. We have $|R_{\hat{1}\hat{0}\hat{1}\hat{0}}|=|R_{\hat{l}\hat{t}\hat{l}\hat{t}}|$. Then the first tidal constraint reduces to
\begin{eqnarray}
|R_{\hat{l}\hat{t}\hat{l}\hat{t}}| &=& \frac{\Delta_1}{4 f \sin^{2}  \theta} |2\,\sin^2 \theta \Delta f \frac {\partial^{2} f}{\partial {l}^{2}}   \\
 &&-\Delta_1 f \sin^2 \theta \frac {\partial f}{\partial l}\frac {\partial K}{\partial l} 
 - K  \sin^2 \theta \left(\frac {\partial f}{\partial \theta}\right)^{2}  \nonumber \\
&&+ f  \sin^{2}\theta  \frac {\partial f }{\partial \theta} \frac {\partial K}{\partial \theta} 
 +  f^{4} K \left(\frac {\partial \omega }{\partial l}\right)^{2} |
\leq (10^{5}km)^{-2} \nonumber .\label{TC1}
\end{eqnarray}

While the lateral constraints are reduced to the study of $|R_{\hat{2}\hat{0}\hat{2}\hat{0}}|\leq (10^{5}km)^{-2}$, and $|R_{\hat{3}\hat{0}\hat{3}\hat{0}}|\leq (10^{5}km)^{-2}$ since our metric is axially symmetric, we have that
\begin{eqnarray}
|R_{\hat{2}\hat{0}\hat{2}\hat{0}}|=\gamma^{2}|R_{\hat{\theta}\hat{t}\hat{\theta}\hat{t}}|
+\gamma^{2}(v^{2}/c)|R_{\hat{\theta}\hat{r}\hat{\theta}\hat{r}}|.
\end{eqnarray}

Now we assume that, for example, the spaceship is at rest at the throat \cite{MT}, this implies $v\rightarrow 0$ and $\gamma\rightarrow 1$. Then $|R_{\hat{2}\hat{0}\hat{2}\hat{0}}|=|R_{\hat{\theta}\hat{t}\hat{\theta}\hat{t}}|$. Then second tidal constrain is given by
\begin{eqnarray}
|R_{\hat{\theta}\hat{t}\hat{\theta}\hat{t}}| &=& \frac{1}{4\Delta^{2}f \sin^{2} \theta}\,|-\Delta\Delta_1 \sin^{2} \theta \left( {\frac {\partial f }{\partial l}}\right)^{2}    \label{TC2}\\
&&+ f   {\Delta_1}    \sin^{2}  \theta {\frac {\partial f }{\partial l}} {\frac {\partial \Delta }{\partial l}}  +2\,  f\Delta \sin^{2}  \theta  {\frac {\partial ^{2} f}{\partial {\theta}^{2}}} \nonumber \\
&& - f \sin^{2}\theta {\frac {\partial f}{\partial \theta}}  {\frac {\partial \Delta }{
\partial \theta}}   + f^{4}K \left( {\frac {\partial \omega }{\partial \theta}}
  \right) ^{2} | \leq (10^{5}km)^{-2}\nonumber
\end{eqnarray}
Finally our last constraint can be written as
\begin{eqnarray}
|R_{\hat{\phi}\hat{t}\hat{\phi}\hat{t}}|&=&\frac{1}{4 f \Delta \sin ^{2} \theta}\,| -  {\Delta_1}\sin^{2} \theta\left( {\frac {\partial f}{\partial l}} \right) ^{2} \nonumber \\
  &&+f  \sin ^{2}\theta {\frac {\partial f}{\partial l}}{\frac {d {\Delta_1}}{dl}} 
   + f^{4}\left( {\frac {\partial \omega }{\partial l}}  \right)^{2} \nonumber \\
  && +2\,f \sin  \theta \cos  \theta  {\frac {\partial f }{\partial \theta}}
- \sin^{2}\theta \left( {\frac {\partial f }{\partial \theta}} \right)^{2} \nonumber	 \\
&&+ \frac{f^4}{\Delta_1} \left( {\frac {\partial \omega }{\partial \theta}} \right) ^{2} |
\leq (10^{5}km)^{-2}.\label{TC3}
\end{eqnarray}

\begin{figure}[ht]
  \centering
  \includegraphics[width=8cm]{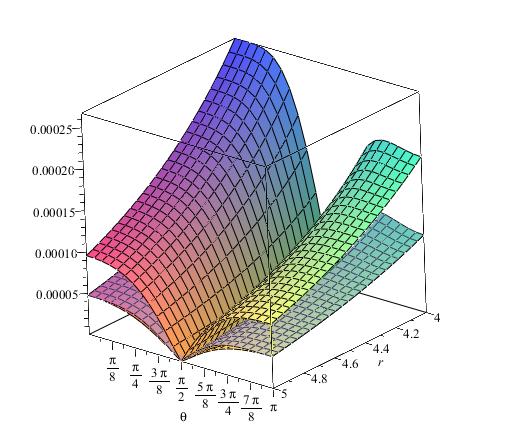}
  \caption{Tidal forces for metric \eqref{eq:solBL}, given by equations (\ref{TC1}), (\ref{TC2}) and (\ref{TC3}). Here $l_1=1$ $l_2=1.5$, $a=0.1$, $k=0.11$. This plot shows the tidal constraints in terms of (\ref{tetrad}).} \label{fig:TC}
\end{figure}

We plot the tidal constraints (\ref{TC1}), (\ref{TC2}) and (\ref{TC3}) in the Fig. \ref{fig:TC}.

\begin{figure}[ht]
  \centering
  \includegraphics[width=8cm]{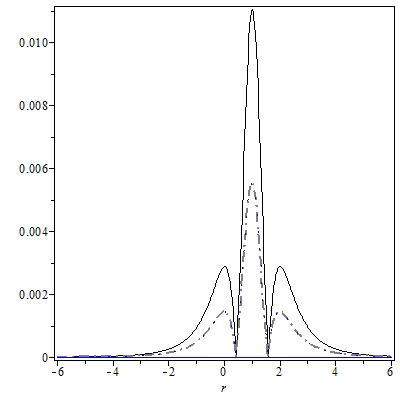}
  \caption{Where $l_1=1$ $l_2=1.5$,
  $a=0.1$,
  $k=0.11$. This plot shows the tidal constraints with $\theta=0$.}\label{fig:TCD}
\end{figure}

We can also force the traveller to approach the throat using the geodesic with $\theta=0$, in that case we have that $\partial w / \partial \theta =0$ and $\partial f / \partial \theta =0$. We can see that trough this geodesic the Riemann tensors simplifies hardly. We plot this result in Fig. \ref{fig:TCD}

\section{Conclusions} \label{sec:Conclusions}

We have analysed metric (\ref{eq:solBL}) and shown that this metric represents a wormhole, which throat is shown in Fig. \ref{fig:Embendding-d}. This metric contains a ring singularity very similar to the Kerr solution, all the invariants of the metric are regular everywhere except on this ring. The throat is found outside of the ring singularity, the mouth of the throat is on a sphere of radius $l=l_1$, around the wormhole. We have shown that the null polar geodesics, that means, geodesics going through the polar line are regular, an observer can go through the throat if the observer trajectory remains on the polar geodesics and contains an energy bigger than $1/2f$, for any values of the free parameters. On this trajectory the tidal forces are very small, therefore this wormhole is traversable. For any other angle the mouth of the wormhole lie on the sphere, but close to the equator, the effect of the wormhole is to repel the test particles. On the equator, the repulsion is infinity and nothing can reach the singularity, even the light is repealed by the wormhole (see also \cite{Matos:2012gj}). Thus, the sphere of radius $l=l_1$ has an effect contrary to the horizon of a black hole, namely, an observer can reach the sphere, goes thought the throat, but this sphere avoids the traveller to observe or to reach the singularity. Of course, the traveller can come back to its original world without much troubles.

\section*{Acknowledgements} We would like to thank
 Dario Nu\~nez for many helpful discussions. The numerical computations were
carried out in the "Laboratorio de Super-C\'omputo
Astrof\'{\i}sico (LaSumA) del Cinvestav", in the UNAM's cluster Kan-Balam and in the cluster Xiuhcoatl from Cinvestav. This work was partially supported by CONACyT M\'exico
under grants CB-2009-01, no. 132400, CB-2011, no. 166212,  and I0101/131/07 C-
234/07 of the Instituto Avanzado de Cosmologia (IAC) collaboration
(http://www.iac.edu.mx/). GM is supported by a CONACYT scholarships.

\end{document}